# A Survey of Fermi Catalog Sources using data from the Milagro Gamma-Ray Observatory

A. J. Smith for the Milagro Collaboration
*University of Maryland, Dept. of Physics, College Park, MD 20742, USA*

The Fermi LAT has released a list of the most significant 205 sources with three months of Fermi data (Bright Source List). The Milagro Gamma-Ray Observatory is sensitive to gamma rays above 100 GeV with a peak sensitivity between 10 and 30 TeV, overlapping and extending the energy range of Fermi. Of the 34 Galactic LAT sources in the field of view of Milagro, 6 are observed with significance greater than 5 sigma and 14 are observed at greater than 3 sigma. Of these 14 sources, 9 are pulsars. Since the VHE emission detected by Milagro is often found to be extended and likely un-pulsed, the VHE component presumably arises from the pulsar winds. Six of the 14 sources have not been previously detected at TeV energies. The details of the Milagro survey will be presented. We will also present the energy spectra of the high-significance detections. Should the full 1-year source list be available prior to the symposium, we will expand our analysis to include the larger Fermi catalog.

## 1. INTRODUCTION

In this paper we report on a recently published analysis by the Milagro collaboration [1]. It was the intention of the authors to extend this analysis through correlation of the Milagro TeV sky survey with the full Fermi catalog, but this analysis could not be completed prior to the deadline for submission of this proceeding, so only the results on the correlation with the Fermi Bright Source List (BSL) are presented.

Though the full Fermi source list has not been released, the Fermi source catalog has been expanded by the publication of a list of 48 pulsars [2]. This list includes 6 northern hemisphere sources that were not in the BSL that we report. We report on these sources. Also presented are preliminarily spectra for the high-significance Milagro detections.

The Milagro gamma-ray observatory has performed the most sensitive survey of 1 to 100 TeV gamma rays from the Northern Hemisphere sky [3,4]. The Milagro data set is ideal for searching for new classes of gamma-ray sources. The 2009 release of the Bright Source List (BSL) by the Fermi collaboration [5] presents such an opportunity by looking for coincidences of > 1 TeV emission with these GeV sources. There are 34 sources in the BSL within Milagro's field of view that are not associated with extragalactic objects. We present a search of the Milagro data for excesses between 1 and 100 TeV coincident with these 34 potential Galactic sources. The analysis presented here differs from previous analyses [1,2,6] by optimizing the event weighting and Gaussian weighting separately in bins of event size (measured with the fraction of channels hit in the instrument). With the improved analysis and an additional year and a half of data, the sensitivity has increased by 15% to 25%, depending on the spectrum of the source.

## 2. THE MILAGRO INSTRUMENT

Milagro was a large-aperture, high-duty-factor instrument sensitive to TeV cosmic rays and gamma rays. The Milagro site is located at 106.68W longitude, 35.88N latitude in northern New Mexico at an elevation of 2650m above sea level. The sensitive area of the detector was comprised of two parts: the central water pond and an 'outrigger' array of water tanks, both instrumented with Photo-Multiplier Tubes (PMTs). The central Milagro pond consisted of two layers of 20 cm-diameter hemispherical Hamamatsu R5912SEL PMTs deployed in a large covered 80x60-meter water pond. 173 water tanks each with a single R5912SEL PMT formed the outrigger array. The median energy for gamma rays detected by Milagro is about 3 TeV, but the peak sensitivity is 20-100 TeV depending on the spectrum due to the fact that the background rejection and the angular resolution improve with energy.

Milagro was the first large area, continuously operating, water Cherenkov detector used for gamma-ray astronomy. Milagro has provided a new and unique look at the Northern sky. Utilizing the wide field-of-view, Milagro has been used to conduct an 8-year survey of the entire overhead sky, covering much of the Northern hemisphere. The major results from Milagro's gamma-ray survey of the sky include the discovery of new TeV gamma-ray sources, discovery of diffuse TeV emission from the plane of the Galaxy and the discovery of diffuse emission from the Cygnus arm of the Galaxy.

## 3. BSL CORRELATON STUDY

We select Fermi-LAT sources in the field of view of Milagro (with $\delta > -5°$) based on their categorization in the BSL. Sources are selected which are confirmed or potential Galactic sources. Sources that are identified as extragalactic are omitted. Sixteen of the selected sources were categorized in the BSL as confirmed pulsars (PSR) and one is a high-mass X-ray binary (HXB). Five sources have a potential association with an SNR, and 12 have no clear association. For each of these 34 sources, we calculate the statistical significance of the Milagro data at the BSL position and estimate the flux or flux limit under the assumption that the emission is from a single point source.

The flux measurements given in Table 1 are derived with a similar approach to [2]. The flux is measured with an assumed spectrum of $E^{-2.6}$ without a cutoff. The dependence of the calculated flux on the true spectrum is minimized when the flux is quoted at the median energy of the hypothesized spectrum. The median energy depends on Declination and varies between 32 and 46 TeV for δ in the range of 10° and 60°. At a Declination of -5°, the median energy of the hypothesized spectrum is 90 TeV. We quote the flux for all sources above 3_ at a representative value of 35 TeV. It should be noted that the median energy used is for the assumed spectrum and not experimentally measured. In particular, a source may in fact cut off before 35 TeV (the Crab for example) and our analysis would still report a flux at 35 TeV.

The results of this search are summarized in Table 1. Of the 34 targets, 14 have a significance greater than 3σ. Six of these are associated with sources or candidates from the first Milagro survey of the Galactic plane [2]. The Crab, MGRO 2019+37, MGRO 1908+06, MGRO 2031+41, and Milagro candidates C3 (likely associated with Geminga) and C4 (likely associated with the Boomerang PWN) are all near LAT GeV sources. In the Milagro data set, the 3σ-5σ observations are fairly marginal because they cannot be convincingly discerned from background when statistical penalties for searching the entire sky are taken into account. However, with LAT points as a trigger for the search, the statistical penalties are reduced. The probability of a single 3σ false-positive in 34 samples of pure background is only 4.4%. The probability of 4 or more excesses at or above 3σ in 34 trials is $1.5 \times 10^{-7}$. It is very likely that most of our 3σ excesses are due to multi-TeV emission. We, therefore, see strong evidence for multi-TeV emission associated with Galactic LAT BSL sources as a class, even if individual sources are not strong enough to definitively distinguish.

There is some contribution to these measurements from the Galactic diffuse emission, but that contribution is small. We can make a conservative estimate by taking the Milagro measurement of the diffuse emission [3] at its highest value, in the inner Galaxy (30° < l < 65°, |b| < 2°). Using this value, we expect $5.3 \times 10^{-17}$ TeV$^{-1}$s$^{-1}$cm$^{-2}$ in a 1° bin at 35 TeV, which is only about 15% contamination for the weakest sources in Table 1. The GALPROP conventional model, for comparison, would only constitute 3% contamination. The contamination is likely lower than suggested by the Milagro measurement because of unresolved sources, such as many of the sources from Table 1. It has even been suggested [7] that most of the Milagro diffuse measurement could be due to unresolved sources. Finally, the Fermi points observed at 3σ in the Milagro data occur near local maxima in the Milagro data. In contrast, the diffuse emission is expected to vary slowly across the Galaxy.

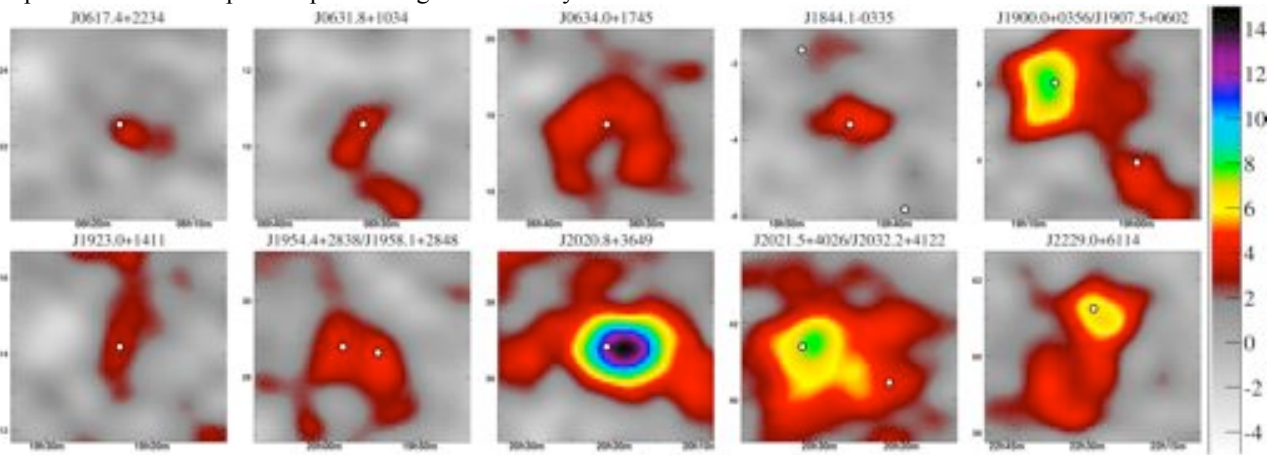

Figure 1: Images of the Fermi >3 excesses identified by Milagro that are coincident with northern hemisphere Fermi BSL sources. The Crab is not shown.

## 4. DISCUSSION OF BSL CORRELATION

From this analysis, it appears quite common for Galactic 100 MeV - 100 GeV sources to have associated multi-TeV emission. This association is notable for pulsars, where 9 of 16 pulsars from the BSL are on our list of likely multi-TeV emitters. The pulsars in the BSL, which have less than 3σ significance in Milagro data tend to lie off the Galactic plane. The pulsars off the plane are typically older, having traveled far from their origin after the kick they received from the initial asymmetric supernova [8]. Of the SNR sources on the list, we see 3 of 5. Interestingly, we see only 2 of the 12 unidentified sources. These unidentified sources may be extragalactic and not visible with this analysis, which was optimized for high-energy emission. Figure 1 shows the regions in the Milagro data around the indicated LAT sources. Eight of the 13 sources are associated with previously reported >TeV sources or candidates:

0FGL J0534.6+2201 is the young Crab Pulsar. Its associated pulsar wind nebula (PWN) is a standard reference source in TeV astronomy.

0FGL J0617.4+2234 is associated with SNR IC443, which is interacting with a nearby large molecular cloud. An associated x-ray feature has been interpreted as a PWN [9], implying the existence of a pulsar, but the no pulsed emission has yet been detected. IC443 was first reported above 1 TeV by MAGIC [10] and later confirmed by VERITAS [11]. The flux reported in Table 1 is somewhat higher than the flux predicted by extrapolating the MAGIC fit, but is roughly consistent after allowing for the extremes of the statistical and systematic errors of the two measurements.

| Name (0FGL) | type | RA (deg) | DEC (deg) | l (deg) | b (deg) | Flux ($\times 10^{-17}$ TeV$^{-1}$ sec$^{-1}$ cm$^{-2}$) | Signif. ($\sigma$'s) | TeV assoc. |
|---|---|---|---|---|---|---|---|---|
| J0007.4+7303 | PSR | 1.85 | 73.06 | 119.69 | 10.47 | < 90.4 | 2.6 | |
| J0030.3+0450 | PSR | 7.60 | 4.85 | 113.11 | -57.62 | < 20.9 | -1.7 | |
| J0240.3+6113 | HXB | 40.09 | 61.23 | 135.66 | 1.07 | < 26.2 | 0.7 | LSI +61 303 |
| J0357.5+3205 | PSR | 59.39 | 32.08 | 162.71 | -16.06 | < 16.5 | -0.1 | |
| J0534.6+2201 | PSR | 83.65 | 22.02 | 184.56 | -5.76 | 162.6 ± 9.4 | 17.2 | Crab |
| J0613.9-0202 | PSR | 93.48 | -2.05 | 210.47 | -9.27 | < 60.0 | -0.0 | |
| J0617.4+2234 | SNR$^a$ | 94.36 | 22.57 | 189.08 | 3.07 | 28.8 ± 9.5 | 3.0 | IC443 |
| J0631.8+1034 | PSR | 97.95 | 10.57 | 201.30 | 0.51 | 47.2 ± 12.9 | 3.7 | |
| J0633.5+0634 | PSR | 98.39 | 6.58 | 205.04 | -0.96 | < 50.2 | 1.4 | |
| J0634.0+1745 | PSR | 98.50 | 17.76 | 195.16 | 4.29 | 37.7 ± 10.7 | 3.5 | MGRO C3 Geminga |
| J0643.2+0858 | | 100.82 | 8.98 | 204.01 | 2.29 | < 30.5 | 0.3 | |
| J1653.4-0200 | | 253.35 | -2.01 | 16.55 | 24.96 | < 51.0 | -0.5 | |
| J1830.3+0617 | | 277.58 | 6.29 | 36.16 | 7.54 | < 32.8 | 0.2 | |
| J1836.2+5924 | PSR | 279.06 | 59.41 | 88.86 | 25.00 | < 14.6 | -0.9 | |
| J1844.1-0335 | | 281.04 | -3.59 | 28.91 | -0.02 | 148.4 ± 34.2 | 4.3 | |
| J1848.6-0138 | | 282.16 | -1.64 | 31.15 | -0.12 | < 91.7 | 1.7 | |
| J1855.9+0126 | SNR$^a$ | 283.99 | 1.44 | 34.72 | -0.35 | < 89.5 | 2.2 | |
| J1900.0+0356 | | 285.01 | 3.95 | 37.42 | -0.11 | 70.7 ± 19.5 | 3.6 | |
| J1907.5+0602 | PSR | 286.89 | 6.03 | 40.14 | -0.82 | 116.7 ± 15.8 | 7.4 | MGRO J1908+06 HESS J1908+063 |
| J1911.0+0905 | SNR$^a$ | 287.76 | 9.09 | 43.25 | -0.18 | < 41.7 | 1.5 | |
| J1923.0+1411 | SNR$^a$ | 290.77 | 14.19 | 49.13 | -0.40 | 39.4 ± 11.5 | 3.4 | HESS J1923+141 |
| J1953.2+3249 | | 298.32 | 32.82 | 68.75 | 2.73 | < 17.0 | 0.0 | |
| J1954.4+2838 | SNR$^a$ | 298.61 | 28.65 | 65.30 | 0.38 | 37.1 ± 8.6 | 4.3 | |
| J1958.1+2848 | PSR | 299.53 | 28.80 | 65.85 | -0.23 | 34.7 ± 8.6 | 4.0 | |
| J2001.0+4352 | | 300.27 | 43.87 | 79.05 | 7.12 | < 12.1 | -0.9 | |
| J2020.8+3649 | | 305.22 | 36.83 | 75.18 | 0.13 | 108.3 ± 8.7 | 12.4 | MGRO J2019+37 |
| J2021.5+4026 | PSR | 305.40 | 40.44 | 78.23 | 2.07 | 35.8 ± 8.5 | 4.2 | |
| J2027.5+3334 | | 306.88 | 33.57 | 73.30 | -2.85 | < 16.0 | -0.2 | |
| J2032.2+4122 | PSR | 308.06 | 41.38 | 80.16 | 0.98 | 63.3 ± 8.3 | 7.6 | TEV 2032+41 MGRO J2031+41 |
| J2055.5+2540 | | 313.89 | 25.67 | 70.66 | -12.47 | < 17.6 | -0.0 | |
| J2110.8+4608 | | 317.70 | 46.14 | 88.26 | -1.35 | < 24.1 | 1.1 | |
| J2214.8+3002 | | 333.70 | 30.05 | 86.91 | -21.66 | < 20.7 | 0.6 | |
| J2229.0+6114 | PSR | 337.26 | 61.24 | 106.64 | 2.96 | 70.9 ± 10.8 | 6.6 | MGRO C4 |
| J2302.9+4443 | | 345.75 | 44.72 | 103.44 | -14.00 | < 13.2 | -0.6 | |

**Table 1** List of Milagro measure fluxes and flux limits for the 34 northern hemisphere galactic sources in the Fermi BSL.

0FGL J0634.0+1745 is the Geminga pulsar. Geminga is a relatively old (342 kyr) but very near (169 pc) pulsar [12,13]. It is the most significant Fermi-LAT source in the northern sky, but emission over 1 TeV has only been reported by Milagro as candidate C3 with too low a significance to be classified as a definitive detection. Milagro observes an emission region that is extended by several degrees as shown in Figure 1. The significance reported in Table 1 has been computed assuming point source emission, but if we instead assume that the source is due to emission from an extended region and convolve a $1^O$ Gaussian with the energy-dependent point spread function, the significance at the location of 0FGL J0634.0+1745 increases to 6.3σ. The local maximum of the Milagro excess is at RA=6h32m28s, Dec=17$^O$22m. Given the high significance, we regard this as a definitive detection of extended emission from Geminga. A spatial Gaussian fit to the data yields a region with a standard deviation of $1.30^O \pm 0.20^O$. For comparison, the analogous fit for the Crab, which is effectively a point source, has a width of $0.6^O$. This suggests that the full width at half maximum of the region of emission in the vicinity of Geminga is $2.6^O + 0.7^O - 0.9^O$, after accounting for the point spread function. The large extent (implying an emission region of some 5 to 10 pc extent) is likely due to the nearness of the source and may arise from a pulsar-driven wind; it is consistent with HESS observation of more distant PWN with an angular size of ~10 pc. This may also explain why the source has not yet been observed by Imaging Atmospheric Cherenkov Telescopes [14].

0FGL J1907.5+0602 is associated with MGRO J1908+06 [2]. This pulsar was discovered by the LAT and is also coincident with AGILE source 1AGL J1908+0613 [15] and EGRET source GEV J1907+557 [16]. The multi-TeV emission was first reported by Milagro. HESS both confirmed the Milagro detection and was also able to identify this source as extended by $0.21^O + 0.07^O - 0.05^O$ [17]. The peak of the Milagro detection occurs at RA=19h6m44s, Dec=5_50m with a 1 sigma error circle of $0.27^O$ and a local peak significance of 8.1σ. The peak of the Milagro emission is .3$^O$ from the pulsar, but consistent with the pulsar's location within the measurement error.

0FGL J1923.0+1411 is associated with SNR G49.2-0.7 (W51) which is in a star-forming region and near molecular clouds. Recently, a >TeV source, HESS J1923+141 (Feinstein et al. 2009), has been detected which is spatially extended and coincident with the Fermi source. 0FGL J2020.8+3649 is associated with MGRO J2019+37. This is the most significant source in the Milagro data set apart from the Crab. The young central pulsar has a period of $10^4$ ms and an estimated age of 17.2 kyr. This source was also detected by AGILE and EGRET. It was AGILE that first identified the GeV pulsations [18] and that discovery was confirmed with Fermi data. The peak of the flux measured by Milagro is at RA=20h18m43s Dec=36$^O$42m with a $0.09^O$ 1-sigma error circle. The position of the excess is _0.3$^O$ from the pulsar.

0FGL J2032.2+4122 is a LAT identified pulsar that is spatially coincident with the HEGRA source J2032+41 [19], MGRO J2031+41, and the MAGIC source J2032+4130 [20]. The Milagro source was reported [2] with an extent of $3^O$, but it appears that the Milagro extended source may be due to two or more overlapping sources with a potential additional diffuse contribution from the highly emissive Cygnus region. The location of the Milagro peak is RA=20h31m43s and Dec=40$^O$40m with a statistical error of $0.3^O$.

0FGL J2229.0+6114 is coincident with the radio pulsar J2229+6114 which has been previously associated [21] with the EGRET source 3EG J2227+6122. The period of this pulsar is 52 ms, its distance is 0.8 kpc [22], and the age is estimated to be 10.5 kyr and E-dot is 2.2 x$10^{37}$ ergs/sec [23]. Milagro detects a 6.6σ excess at the position of the pulsar and a local maximum of 6.8σ. The peak of the Milagro excess is RA=22h28m44s Dec=61$^O$10m with a statistical position error of $0.165^O$. This source was reported as candidate C4 by Milagro in [2]. With the additional data and improved analysis presented here, this source is elevated to a high-confidence detection. Milagro also identifies this source as clearly not a point source, with a long extension to the south.

The remaining five objects with greater than 3σ excess in the Milagro data have not been previously detected above 1 TeV energies:

0FGL J0631.8+1034 is the radio pulsar J0631+10 [24]. This pulsar has a period of 288 ms and an estimated age of 43.6 kyr, a distance of 6.55 kpc and E-dot of $1.7 \times 10^{35}$ erg/s [28]. The VERITAS upper-limit for this region is 1.3% of the Crab [25].

0FGL J1844.1-0335 is unassociated with any known source. It is an interesting source because it occurs at a Declination at the edge of Milagro's sensitivity and, if the Milagro observation is real, it is extremely bright above 1 TeV. It is in the region of the Galactic plane surveyed by HESS (Aharonian et al. 2006) but was not detected. To account for the HESS non-detection, the source would have to be extended or have a very hard spectrum extending to high energy. 0FGL J1900.0+0356 has no known associations. 0FGL J1954.4+2838 is coincident with SNR G65.1+0.6 which has been associated wit PSR 1957+2831 (Tian & Leahy 2006). 0FGL J1958.1+2848 is a LAT-discovered pulsar that is associated with the EGRET source 3EG J1958+2909 [26]. 0FGL J2021.5+4026 is a LAT-discovered pulsar that is coincident with the gamma-Cygni SNR. This source is located in the Cygnus region that is detected by Milagro as having a broad extended excess.

The relationship between the Fermi and Milagro source fluxes and upper limits for these 34 sources is shown in Figure 3. The BSL values for the integral flux are shown with the Milagro measurements of the differential flux at 35 TeV. The 35 TeV fluxes are roughly correlated with the measurements between 100 MeV and 100 GeV but the correlation is not strong, with a correlation coefficient of the 3_ points in log space of only 0.2. One possible explanation for the pulsa variation is that the pulsed emission is expected to be beamed (and thus viewing-angle dependent) and the unpulsed multi-TeV emission is likely unbeamed (Gaensler & Slane 2006). The spectrum that connects the Milagro flux to the Fermi flux is universally softer than 2.0 and closer to 2.3, depending on the source.

We have found that pulsars and/or their associated PWN dominate the population of Fermi sources observed above $3\sigma$ by Milagro. Of the 4 high-confidence Milagro detections associated with pulsars of known periodicity and distance, 3 (namely J0534.6+2201, J0634.0+1745, and J2229.0+6114) have E-dot/$d^2$ above $10^{35}$ ergs s$^{-1}$ kpc$^{-2}$ where E-dot is the spin down luminosity and d is the distance to the pulsar. The distance on the fourth (J2020.8+3649) is uncertain. Using the 3-4 kpc distance implied by x-ray measurements [27] rather than the 12 kpc measurement implied by the pulsar dispersion measurement, it too has E-dot/$d^2$ above $10^{35}$ ergs s$^{-1}$ kpc$^{-2}$. HESS reported a similar association with high E-dot/$d^2$ pulsars [28]. Since the pulsed emission is beamed and the PWN is not, all high E-dot/$d^2$ pulsars are possible candidates for multi-TeV emission. We have searched the ATNF pulsar database [29] for northern-hemisphere pulsars with a high E/d2, which were not reported in the Fermi BSL. Of the 25 highest E/$d^2$ pulsars, there are 10 in the northern hemisphere and 5 not identified as GeV sources by Fermi. These 5 are J0205+6449, J0659+1414, J1930+1852, J1913+1011 and J1740+1000. Of these, the largest statistical significance was 3.3 standard deviations (PSR J1930+1852), not significant enough to claim this as new source of multi-TeV gamma rays (though follow-up observations are warranted).

Finally, it is interesting to note that of the sources published in the Milagro survey of the Galactic plane [2], all 4 sources and two of the 4 source candidates are now strongly associated with pulsars, suggesting that most of the Milagro sources are multi-TeV PWN. We also note the high efficiency with which MeV to GeV pulsars are observed above 1 TeV, and a qualitative picture is emerging where the typical Galactic multi-TeV source is a PWN associated with a MeV to GeV pulsar.

## 5. PULSAR CATALOG CORRELATION

The Fermi collaboration has released a catalog of pulsars. 46 GeV identified pulsars are included, with 23 of the sources located within the field of view for Milagro. There is considerable overlap between the BSL and the pulsar catalog. Only 7 of the 23 northern hemisphere sources were identified as part of the BSL. These 7 sources are listed in table 2 below.

One of these pulsars, J2238+59, was announced by Fermi prior to the submission, but before the publication of the Milagro BSL correlation paper. Though this source was not included in the broader analysis, it was mentioned in a footnote. This source is only a ~2 deg from the bright pulsar J2229+6114 where Milagro detects a high-significance signal that is extended to the south (see figure 1). This excess is likely due to a second object, since J2238+59 lies directly at the center of the region of excess.

Aside from J2238+39, which is detected with a $4.7\sigma$ excess by Milagro, none of the other 6 Fermi pulsars are observed with greater than $3\sigma$ excess in Milagro.

| PSR | lon (deg) | lat (deg) | P-dot | E-dot $10^{34}$ erg/s | d (kpc) | E-dot/$d^2$ $10^{34}$ erg/s/kpc$^2$ | Energy Flux ergs/cm$^2$/s | Milagro ($\sigma$'s) ($\sigma$'s) |
|---|---|---|---|---|---|---|---|---|
| J0205+6449 | 130.7 | 3.1 | 194 | 2700 | 2.9 | 321.0 | 6.6 | -1.0 |
| J0218+4232 | 139.5 | -17.5 | $7.7 \times 10^{-6}$ | 24 | 3.25 | 2.3 | 3.6 | 2.9 |
| J0248+6021 | 137.0 | 0.4 | 55.1 | 21 | 2-9 | - | 3.07 | 1.54 |
| J0659+1414 | 201.1 | 8.3 | 55.0 | 3.8 | 0.288 | 45.8 | 3.2 | 0.8 |
| J0751+1807 | 202.7 | 21.1 | $6.0 \times 10^{-6}$ | 0.6 | 0.60 | 1.7 | 1.1 | -0.4 |
| J2043+2740 | 70.6 | -9.2 | 1.3 | 5.6 | 1.80 | 1.7 | 1.6 | -0.9 |
| J2238+5900 | 106.4 | 0.5 | 98.6 | 90 | - | - | 5.4 | 4.7 |

Table 2 List of Fermi GeV Pulsars observable by Milagro that are were not included in the BSL.

## 6. ENERGY SPECTRA

To perform an energy spectrum fit, the data was separated into 9 energy bins. The energy of each event was estimated using a simple sum of the fraction of PMTs hit in the top layer and the fraction of outrigger PMTs hit. This parameter, denoted $\mathcal{F}$, ranges from 0.2 to 2. Rather than converting these to energy and assigning an energy value to each event, we perform the fit to the energy spectrum in the space of $\mathcal{F}$. This avoids systematic errors that can arise from a complex energy resolution function, which is potentially important since the energy resolution of Milagro is poorer than satellite experiments and IACTs, which can integrate the total energy of the shower rather than sampling the longitudinal tail. The energy resolution for Milagro is generally log-normal and ranges from ~0.5 decades at a few TeV to ~0.2 decades at 100 TeV. The fit is performed using the hypothetical spectrum, $dN/dE = I_0 E^{\lambda} \exp(-E/E_{cut})$. For each value of $I_0$, $E_{cut}$ and $\lambda$ the predicted $\mathcal{F}$ distribution is computed using a detailed Monte Carlo simulation. $\chi^2$ is computed by comparing the predicted and measured $\mathcal{F}$ distributions.

The blue and yellow regions in figures 2-5 show the 1-σ and 2-σ confidence intervals for energy spectrum of 4 high significance sources detected by Milagro. Three of these sources, The Crab, PSR J1908+06 and PSR J2227+6114 have also been detected by IACTs. For these sources the HESS/VERITAS fit results and spectra have been overlaid.

The spectrum of the Crab shows excellent agreement between Milagro and HESS. Only at the HESS data at the highest energy strays outside the Milagro confidence interval. We regard this agreement as a verification of the energy analysis of Milagro and the simulation of the sensitivity of the Milagro instrument. Note also that since Milagro observes the entire overhead sky at all times and this analysis is based on 3 years of data, any systematic deficiencies that are present in the Milagro data at one point in the sky should be present at all other points. This is not in general true for pointed instruments, which collect data from different source at different times and under differing conditions, whereas data in Milagro is collected concurrently for the entire sky.

Figure 3 shows the spectrum for PSR J1907.5+0602. After the discovery of this TeV source by Milagro, HESS and VERITAS confirmed it. The spectrum derived from the HESS observations has been overlaid. Unlike the case with the Crab, there is an apparent discrepancy between the spectrum measured by Milagro and that of HESS, however since the error bars are large the discrepancy is probably only between 2σ and 3σ. While this could be simply due to a statistical fluctuation, one other consideration is that this source is not point-like, so the superior angular resolution of HESS isolates the source core, whereas the poorer angular resolution of Milagro integrates more of the diffuse lateral tails. In such as case, one would expect that the Milagro flux would be higher, which is what we observe. HESS fits the spectrum of this source to a hard power-law (λ=2.1) and finds no evidence of a cut off. However, their energy reach is limited to <20 TeV. Milagro, likewise, cannot definitively detect a cutoff from this source. Since a soft spectrum with λ~3 and a hard spectrum with a cutoff both yield acceptable values of $\chi^2$. We can however use the conclusions of HESS to constrain the Milagro data. We do this by fixing λ at 2.1. When this is done, the optimal fit gives $E_{cut}$=14 TeV and fits with no cutoff are ruled out. The 1-σ confidence interval for the cutoff is [10 TeV, 40 TeV] and the 2-σ confidence interval includes [7 TeV, 56 TeV]. The spectrum under this assumption is definitively constrained to cut off.

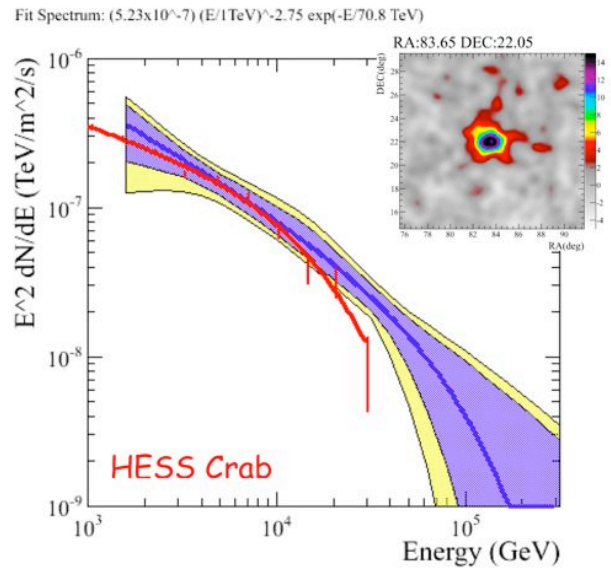

Figure 2 Confidence intervals (1-sigma in blue and 2-sigma in yellow) for fit to the spectrum of the Crab.

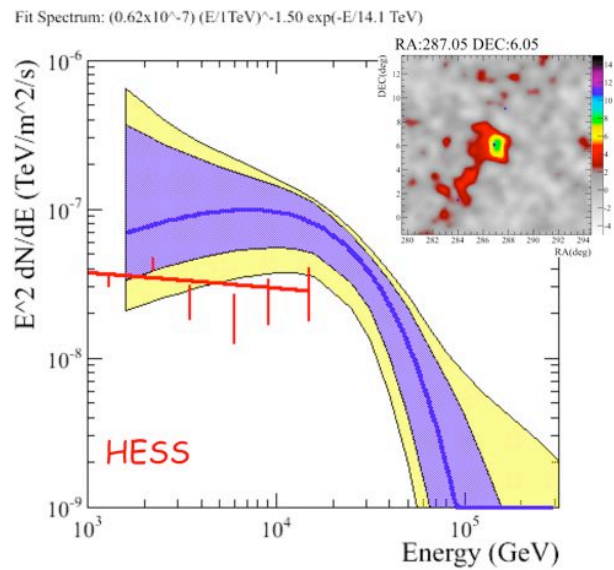

Figure 3 Confidence intervals (1-sigma in blue and 2-sigma in yellow) for fit to the spectrum of the PSR J1908+06. Data and fit results from HESS have been overlaid.

Figure 4 shows the spectrum measure for Milagro source MGRO J2019+37, which is coincident with Fermi PSR J2020.8+3649. This high significance (12.4σ) detection has not been confirmed by other TeV observations. As is the case with MGRO J1908+06, though a hard spectrum with a cutoff is favored, a soft power-law spectrum with no cutoff also has an acceptable $\chi^2$.

Figure 5 shows the spectrum for PSR J2229+6114. Milagro detects this source at 6.6σ. This source, as the others can be adequately fit to either a soft spectrum with no cutoff or a hard spectrum with a cutoff at or above 10 TeV. This source was also reported by VERITAS [30]. The spectrum reported by VERITAS is shown on the figure and is consistent with the Milagro measured spectrum with errors. The spectral index reported by VERITAS has a sufficiently large error that, unlike the case of PSR J1908+06, we cannot use the IACT spectral index measurement to constrain the Milagro fit and definitively rule in or out the presence of a high energy cutoff.

Note that the sensitivity of Milagro to an $E^{-2}$ spectrum has peak sensitivity near 100 TeV though none the TeV sources detected by Milagro show evidence of 100 TeV emission, nor is there evidence in the Milagro data of emission at 100 TeV from any other source. This observation rules out the existence of any source with flux greater than about $10^{-16}$ particles/cm$^2$/s above 100 TeV. This result is likely to constrain theories of VHE particle acceleration and may also have consequences for VHE neutrino detection.

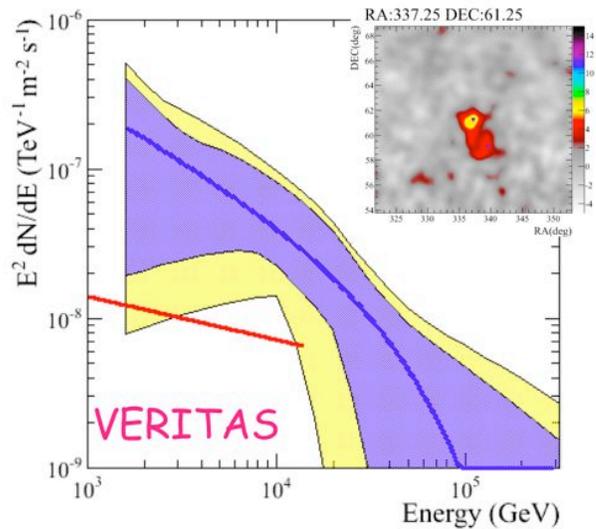

Figure 5 Spectrum of PSR 2229+6114. The Spectrum measured by VERITAS has been overlaid (red line).

## Acknowledgments

We gratefully acknowledge Scott Delay and Michael Schneider for their dedicated efforts in the construction and maintenance of the Milagro experiment. This work has been supported by the

National Science Foundation (under grants PHY-0245234, -0302000, -0400424, -0504201,-0601080, and ATM-0002744), the US Department of Energy (Office of High-Energy Physics and Office of Nuclear Physics), Los Alamos National Laboratory, the University of California, and the Institute of Geophysics and Planetary Physics.

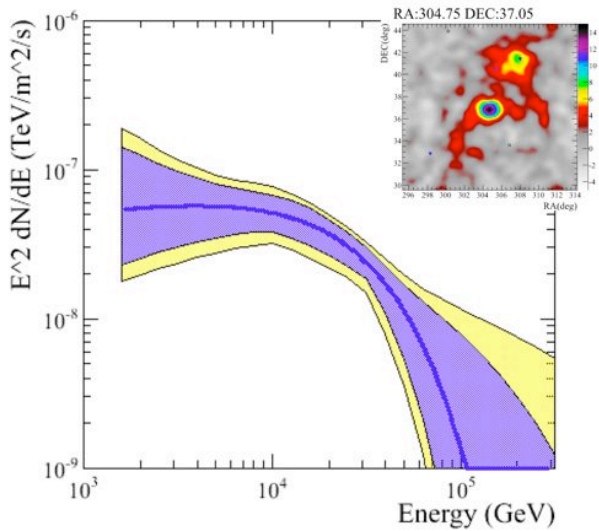

Figure 4 Spectrum of MGRO J2019+37